%% file: vonhippel.tex
\documentclass[graybox,vecphys]{svmult}

\usepackage{type1cm}        
\usepackage{makeidx}         
\usepackage{graphicx}        
\usepackage{multicol}        
\usepackage[bottom]{footmisc}

\usepackage{newtxtext}       %
\usepackage{newtxmath}       

\usepackage{hyperref}

\makeindex             

\begin{document}

\title*{Direct Integration for Multi-leg Amplitudes: Tips,
Tricks, and When They Fail}
\author{Jacob L. Bourjaily, Yang-Hui He, Andrew J. McLeod, Marcus Spradlin, Cristian Vergu, Matthias Volk, Matt von Hippel, and Matthias Wilhelm}
\authorrunning{Matt von Hippel et al.}
\institute{Jacob L. Bourjaily \at Niels Bohr Institute, University of Copenhagen, Blegdamsvej 17, 2100 Copenhagen, Denmark \& Institute for Gravitation and the Cosmos, Department of Physics, Pennsylvania State University, University Park, PA 16892, USA, \email{bourjaily@psu.edu} \and Yang-Hui He \at Department of Mathematics, City, University of London, London EC1V0HB, UK \& Merton College, University of Oxford, OX1 4JD, UK \& School of Physics, NanKai University, Tianjin, 300071, P.R. China, \email{hey@maths.ox.ac.uk} \and Andrew J. McLeod \at Niels Bohr Institute, University of Copenhagen, Blegdamsvej 17, 2100 Copenhagen, Denmark, \email{amcleod@nbi.ku.dk} \and Marcus Spradlin \at Department of Physics, Brown University, Providence, RI 02912, USA, \email{marcus_spradlin@brown.edu} \and Cristian Vergu \at Niels Bohr Institute, University of Copenhagen, Blegdamsvej 17, 2100 Copenhagen, Denmark, \email{c.vergu@nbi.ku.dk}\and Matthias Volk \at Niels Bohr Institute, University of Copenhagen, Blegdamsvej 17, 2100 Copenhagen, Denmark, \email{mvolk@nbi.ku.dk} \and Matt von Hippel \at Niels Bohr Institute, University of Copenhagen, Blegdamsvej 17, 2100 Copenhagen, Denmark, \email{mvonhippel@nbi.ku.dk} \and Matthias Wilhelm \at Niels Bohr Institute, University of Copenhagen, Blegdamsvej 17, 2100 Copenhagen, Denmark, \email{matthias.wilhelm@nbi.ku.dk}}
%
%
\maketitle

\abstract*{Direct hyperlogarithmic integration offers a strong alternative to differential equation methods for Feynman integration, particularly for multi-particle diagrams. We review a variety of results by the authors in which this method, employed with some care, can compute diagrams of up to eight particles and four loops. We also highlight situations in which this method fails due to an algebraic obstruction. In a large number of cases the obstruction can be associated with a Calabi-Yau manifold.}

\abstract{Direct hyperlogarithmic integration offers a strong alternative to differential equation methods for Feynman integration, particularly for multi-particle diagrams. We review a variety of results by the authors in which this method, employed with some care, can compute diagrams of up to eight particles and four loops. We also highlight situations in which this method fails due to an algebraic obstruction. In a large number of cases the obstruction can be associated with a Calabi-Yau manifold.}

\section{Introduction}
\label{sec:1}
Several methods are available for evaluating Feynman integrals in terms of hyperlogarithms. Of these, direct hyperlogarithmic integration is perhaps surprisingly a bit of a dark horse. The method, which in rough outline consists of partial-fractioning rational functions and re-expressing hyperlogarithms in the integration variable, applying the definition of the hyperlogarithm, and careful treatment of boundary values~\cite{Brown:2009ta,Panzer:2015ida}, has been implemented in computer packages~\cite{Panzer:2014caa,Bogner:2014mha}, but remains less popular than differential equation methods~\cite{Kotikov:1990kg,Remiddi:1997ny,MullerStach:2012mp,Henn:2013pwa,Lee:2014ioa,Meyer:2016slj} or methods based on the Mellin-Barnes representation~\cite{Gluza:2007rt,Blumlein:2014maa,Ochman:2015fho}.

In part, this lack of popularity is due to the method's dependence on linear reducibility. At each integration step, it must be possible to express the integrand in terms of rational functions and hyperlogarithms in the integration parameter. If there is no integration order such that this is possible then we say that the integral is not linearly reducible. This can happen, for example, if partial-fractioning in a previous integration step gives rise to irreducible algebraic roots in a later integration variable. If this happens then direct hyperlogarithmic integration is obstructed.

Despite this potential for obstruction, direct hyperlogarithmic integration has several advantages. Differential equation and Mellin-Barnes methods both have difficulty in problems with a large number of scales. In contrast, provided linear reducibility is preserved the number of scales has little impact on the difficulty of direct hyperlogarithmic integration. As such, it is particularly well-suited for multi-leg scattering amplitudes. These are especially relevant in the context of planar $\mathcal{N}=4$ super Yang-Mills, where amplitudes with five particles and fewer are fully captured by the BDS ansatz~\cite{Bern:2005iz}.

In this talk, we present several direct integrations of multi-leg Feynman integrals by the authors, mostly in the context of planar $\mathcal{N}=4$ super Yang-Mills~\cite{Bourjaily:2017bsb,Bourjaily:2018ycu,Bourjaily:2018aeq,Bourjaily:2018yfy,Bourjaily:2019hmc,Bourjaily:2019igt,Bourjaily:2019vby}. We observe that obstructions to linear reducibility can be postponed or avoided entirely via a variety of techniques and best-practices. In some cases, integration can be performed to completion, resulting in an expression in terms of hyperlogarithms with rational arguments~\cite{Bourjaily:2018aeq}. In others, integration can still be completed, but the resulting hyperlogarithms depend manifestly on algebraic roots in the kinematics. In several cases presented here, it is possible to show that this dependence is spurious, and the singularities of the result are in fact all rational in the kinematics~\cite{Bourjaily:2019igt,Bourjaily:2019vby}. In still other cases, integration cannot be completed, and there is unavoidable algebraic dependence on the integration parameters at an intermediate stage. These cases have an intriguing commonality of structure: in each case, the obstructions to direct integration can be characterized in terms of Calabi-Yaus~\cite{Bourjaily:2017bsb,Bourjaily:2018ycu,Bourjaily:2018yfy,Bourjaily:2019hmc}.

\section{Tips and Tricks for a Rational Result}
\label{sec:2}
As an illustrative example, consider the following direct integration, which should be thought of as the first step in a longer calculation:
\begin{eqnarray}
\int_0^\infty  \frac{d \alpha}{\alpha^2+ 2 f \alpha + g} &=&
 \int_0^\infty \frac{d\alpha}{2 \sqrt{f^2- g}}\left( \frac{1}{\alpha + f - \sqrt{f^2- g}} - \frac{1}{\alpha + f + \sqrt{f^2- g}} \right)\\
&=& \frac{1}{2 \sqrt{f^2- g}} \log \left( \frac{f+\sqrt{f^2- g}}{f-\sqrt{f^2- g}} \right) \,.
\end{eqnarray}

If $\sqrt{f^2- g}$ happens to be a perfect square, the result will contain no irreducible roots and direct integration can proceed happily. If it is not, direct integration is obstructed. Our goal then is to avoid obstructions of this kind. Sometimes this can be accomplished simply by choosing a different order of variables in which to integrate: these orders can be found systematically via compatibility graph reduction~\cite{Brown:2008um,Brown:2009ta,Bogner:2013tia,Panzer:2015ida}. In other cases there is no integration order free of these obstructions and they must be dealt with in another way. Sometimes this can be done via a change of variables which rationalizes the square root, which have been studied systematically in ref.~\cite{Besier:2018jen}. 

These systematic methods are often useful, but they do not suffice to avoid every possible obstruction. Going beyond them requires re-thinking our starting assumptions. If we begin with a different representation for the integrand, integrals can be linearly reducible that were not in the original parametrization. This freedom of reparametrization is much less understood. In the following we discuss a few approaches which have been particularly helpful in our calculations of multi-leg diagrams, in which a different framing of a problem can make a seemingly obstructed integral reducible.

\subsection{Loop-by-Loop Parametrization}
\label{subsec:loopbyloop}

It is a widely believed conjecture that, for $L$-loop Feynman integrals in four dimensions, any hyperlogarithms that appear are of maximum weight $2L$. With this in mind, one would expect it to be possible to write any such integral as a $2L$-fold integral over a rational function. This is not typically true of the standard Feynman parameter representation, which has a variable for each propagator, potentially leading to many more than $2L$ integration variables. In practice, this dependence on extra variables can obscure linear reducibility. Heuristically, the more integrations need to be performed the greater the chance that one will introduce a spurious algebraic root. As such, it is wise to begin with a representation depending on as few integration variables as possible, preferably only $2L$.

A method that can achieve this goal in many cases, and approach it in others, is loop-by-loop Feynman parametrization. This method is in some ways analogous to the loop-by-loop approach to the Baikov representation~\cite{Frellesvig:2017aai}, but less general: it is particularly applicable to planar diagrams with massless propagators. In these cases it is often possible to Feynman parametrize one loop at a time, treating the other loop momenta as external. As the diagram is planar, it can be written so that only one propagator depends on a given one of the remaining loop momenta. Then after Feynman parametrizing the first loop, one can integrate in the Feynman parameter corresponding to that propagator, isolating the dependence on the next loop momentum in a ``propagator-like'' form. This then allows the next loop to be integrated in the same way, while at the same time reducing the number of integration parameters in the final result. This method was used more or less straightforwardly to obtain $2L$- or $2L+1$-parameter integrands for $L$-loop diagrams in refs.~\cite{Bourjaily:2017bsb,Bourjaily:2018ycu,Bourjaily:2018aeq,Bourjaily:2019igt,Bourjaily:2019vby}. It was discussed in a bit more detail in ref.~\cite{Bourjaily:2019jrk}. More complicated cases requiring more involved changes of variables were considered in ref.~\cite{Bourjaily:2019hmc}, including a six-parameter representation of a three-loop integral and a nine-parameter representation of a four-loop integral.

\subsection{Momentum Twistors}

Much as it is wise to use as few integration parameters as possible, it is also wise to use as few kinematic parameters as possible. For planar integrals, a particularly natural way to do so is by employing momentum twistor space~\cite{Hodges:2009hk}. To define this space for an $n$-point diagram, we can first consider the dual space defined by dual $x$-coordinates $p_a=x_{a+1}-x_a$, with $x_{n+1}=x_1$. These coordinates automatically enforce momentum conservation. For massless external momenta, we then have the additional constraint that $(x_{a+1}-x_a)^2=0$, so the dual points are light-like separated.\footnote{For massive external momenta, we can represent each in terms of a pair of massless external momenta, so this discussion still applies.} To make this manifest, we go to momentum twistor space, in which each dual point $x_a$ is assigned to a line $\textrm{span}\{z_{a-1},z_a\}$ in $\mathbb{P}^3$. 

In addition to being a natural minimal set of kinematic parameters, momentum twistors also have an additional advantage: they rationalize many of the kinematic square roots that would otherwise occur in scattering amplitudes. Construct the following $n\times n$ matrix of kinematic invariants:
\begin{equation}
G=\{G^a_b=(x_a-x_b)^2\}\,.
\end{equation}
In momentum twistors, the invariant $(x_a-x_b)^2$ is proportional to the determinant $\textrm{det}\{z_{a-1}, z_a, z_{b-1}, z_b\}$. Thus, this matrix is linear in each of the momentum twistors. One can represent each pair $z_{a-1}\wedge z_a$ and $z_{b-1}\wedge z_b$ as a six-component vector, which shows that this matrix is in fact a Gramian matrix, with rank at most six. This means that all $7\times 7$ minors of the matrix should vanish. Expressed in terms of $(x_a-x_b)^2$ or conformal cross-ratios, this will give rise to relations which are quadratic in each such variable, with solutions involving algebraic roots of $7\times 7$ determinants of $G$. These appear quite generically when attempting direct integration of seven- and higher-point amplitudes in $x_a$ variables or in cross-ratios. Because $G$ is manifestly linear in the pairs $z_{a-1}\wedge z_a$ and $z_{b-1}\wedge z_b$, in these variables these algebraic roots are rationalized. Thus momentum twistors serve to rationalize a particularly common class of algebraic roots, permitting direct integration smoothly in more cases.

The combination of loop-by-loop parametrization and momentum twistors is already quite powerful. In ref.~\cite{Bourjaily:2018aeq}, these methods sufficed to compute hyperlogarithmic representations for several classes of integrals, including the six-point ``double penta-ladder'' integrals of ref.~\cite{Caron-Huot:2018dsv}, seven-point integrals referred to as ``Heptagon A'' and ``Heptagon B'', and an eight-point family of integrals referred to as ``Octagon A'', all through four loops. Another eight-point integral in ref.~\cite{Bourjaily:2018aeq}, referred to there as ``Octagon B'', could be computed at two loops through the same methods, resulting in hyperlogarithms which depend on an algebraic square root, arising from a kinematic configuration related to the four-mass box which will be discussed in section \ref{sec:3}. Finally, of the sixteen two-loop six-point integrals computed in ref.~\cite{Bourjaily:2019jrk} and the five seven-point two-loop integrals computed in ref.~\cite{Bourjaily:2019vby}, all but one were computed using essentially the above methods, augmented by a few minor tricks.

\subsection{Splitting the Integration Path}

While loop-by-loop Feynman parametrization and momentum twistors make many planar integrals linearly reducible, they do not fix all obstructions to linear reducibility. In cases when they fail, it is sometimes possible to perform direct integration regardless, via the expedient of splitting the integration path.

There are two distinct versions of this trick, corresponding to two distinct situations. In the first, an integrand may be linearly irreducible due to the occurrence of two polynomials (either in denominators of rational functions or singularities of hyperlogarithms) that do not have a common order in which they can be integrated. In such a situation, it is sometimes the case that the terms containing these polynomials can be separated: the integrand can be written as a sum of terms depending on one polynomial, terms depending on the other, and terms depending on neither. In these situations one can then split the integration, integrating one set of terms with one integration order and the other set of terms with another, so that each set of terms continues to be linearly reducible.

In the second situation, the integrand already contains algebraic roots. A single square root of a quadratic polynomial can always be rationalized via a rational change of variables. If there are multiple such square roots then simultaneously rationalizing them is a nontrivial task. The methods of ref.~\cite{Besier:2018jen} provide criteria to distinguish cases which can be rationalized from cases which cannot. However, it may not always be necessary to simultaneously rationalize all of the algebraic roots occurring in an integrand. This is because these roots, much like the mutually incompatible polynomials of the first situation, may appear only in separate terms of the integrand. Then by separating the integrand into terms each containing only one distinct square root of a quadratic polynomial, it is possible to group the integrand into pieces that can individually be rationalized by distinct changes of variables, thus allowing integration to continue.

The first of these situations held in the fifth integral considered in ref.~\cite{Bourjaily:2019vby}, which was successfully integrated using this method. The integrals considered in ref.~\cite{Bourjaily:2019igt} were more complicated: these required both methods at different stages of the calculation, with the integrand first split into two pieces that integrate rationally along different orders and then split again into pieces that could be rationalized by distinct changes of variables.

\section{Kinematic Square Roots at Symbol Level}
\label{sec:3}

The methods described in the preceding section allowed in several cases for integration to proceed all the way to a hyperlogarithmic expression. However, these expressions themselves may be expressed in terms of algebraic roots in kinematic parameters, even when written with rational parametrizations of momentum twistors.

In general, one expects some algebraic roots that appear in this way to be spurious, artifacts of our integration procedure that are not required to express the integral, while others may be unavoidable. The latter should be ``physically meaningful'' in some sense: it should be possible to characterize them in terms of Landau singularities~\cite{Landau:1959fi}, for example. In the case of planar diagrams with massless internal lines, these roots can originate from Gramian determinants smaller than the $6\times 6$ determinants that momentum twistors rationalize, such as $4\times 4$ Gramian determinants. These are the origin of the square root appearing in the four-mass box~\cite{Hodges:boxInt,Hodges:2010kq,Mason:2010pg}, which indeed is not rationalized with momentum twistors alone. This same kinematic behavior is responsible for the square root observed in the integral referred to as Octagon B in ref.~\cite{Bourjaily:2018aeq}, as well as for the square roots that were expected for the integrals considered in ref.~\cite{Bourjaily:2019igt}.

Ideally we would like to represent an integral using whichever algebraic roots are necessary and no more, eliminating any spurious arguments in our hyperlogarithms. If our hyperlogarithms depended on rational arguments we could do this by going to a uniquely specifiable basis called a fibration basis~\cite{Brown:2009qja,Panzer:2015ida}. However, this is not possible when the arguments of the hyperlogarithms contain algebraic roots. When a root can be rationalized by another change of variables it is sometimes possible to go to a fibration basis after doing so and then transform back to show that the root does not contribute. This method was used in ref.~\cite{Bourjaily:2019jrk}. When this is not possible, sometimes one can propose a plausible ansatz of hyperlogarithms without algebraic roots, then match series expansions at a sufficient depth to be convincing. If neither of those are possible, then one generally cannot find an explicit hyperlogarithmic form without roots, but one may still be able to identify some roots as spurious at the level of the symbol~\cite{Goncharov:2010jf}.

Before discussing this, we should clear up some common misunderstandings regarding the symbol map. In particular, there are two slogans that are often repeated: \textbf{``The symbol trivializes all identities''} and \textbf{``The symbol of a constant vanishes''}. Both of these slogans are useful in the proper context, but neither is strictly true.

The symbol results from maximal iteration of the coaction on a hyperlogarithm, resulting in a tensor product of logarithms. If these are all logarithms of rational  numbers and functions then the symbol will indeed trivialize all identities: one simply needs to factor the argument of each logarithm and expand. However, if any of the numbers or functions involved are algebraic then this procedure will not typically be unique, and thus will not trivialize all identities. This is because algebraic extensions of the rationals are in general not unique factorization domains.

To give a simple example, consider the integers extended by $\sqrt{-5}$. The number nine can be factorized in this ring in two different ways:
\begin{equation}
9=3\times 3=(2+\sqrt{-5})\times(2-\sqrt{-5})\,.
\label{eq:nonunique}
\end{equation}
As neither $3$, $(2+\sqrt{-5})$, or $(2-\sqrt{-5})$ can be factorized further in this ring (more precisely they are irreducible elements), this shows that there are numbers which cannot be uniquely factorized.

In order to make the symbol useful in the presence of algebraic roots, then, we need to find a basis of symbol letters that is truly linearly independent, preferably where as few letters as possible are algebraic. If there are few enough algebraic letters this can be done by inspection, or almost as easily. This was the case for the fifth integral considered in ref.~\cite{Bourjaily:2019vby}: its symbol depended on only one algebraic root, which appeared in 22 distinct letters. It was reasonably straightforward to find relations between these letters, expressing them in a basis of just five algebraic letters (as well as assorted rational letters). When the symbol was expanded in this basis all dependence on the five remaining algebraic letters dropped out, resulting in a purely rational symbol alphabet. 

If there are many algebraic letters, especially of higher degree, then a more systematic approach is desirable. We will discuss such an approach below, employing software implementations of algebraic extensions of the integers. In order to do this we will have to consider constant symbol letters, which brings us to the second misleading slogan, the claim that the symbol of a constant vanishes. This slogan may seem plausible to readers used to calculations in planar $\mathcal{N}=4$ super Yang-Mills, where the constants of interest at well-behaved kinematic points are typically multiple zeta values. As each term in the symbol of a multiple zeta value contains at least one entry equal to one, and $\ln 1=0$, it is true that the symbol of a multiple zeta value vanishes. However, we do not need to choose a well-behaved kinematic point. Choosing instead a generic kinematic point results in a non-vanishing constant symbol, with all of the properties that make symbols useful for non-constant functions. 

With the above in mind, we integrated the eight-point integrals investigated in ref.~\cite{Bourjaily:2019igt} at a particular, generic kinematic point, computed their symbols. These symbols were initially particularly complicated: one of the two integrals considered had a symbol with 8,367,616 terms in 2,024 letters, while the other had 9,941,483 terms in 2,156 letters. These initially involved very complicated algebraic numbers, in some cases up to degree 16. The most complicated letters had a common structure: they were of the form $\rho-\sigma$, where $\sigma$ was a root of a fourth-order polynomial and $\rho$ was a linear combination of at most two square roots and an integer. By grouping these letters according to the roots appearing in $\rho$ and $\sigma$, it was possible to search for combinations that do not involve higher than square roots. This search was accomplished with the use of \texttt{SageMath}~\cite{sagemath}, in particular its \texttt{Pari}~\cite{PARI2} functionality. The relations found in this way were sufficient to remove all higher roots from the symbols, leaving letters that were linear combinations of at most two square roots.

These letters still satisfy many nontrivial relations. To find them, we employ factorization in prime ideals. We sketch the method below:

The ideal generated by a number is defined as the set of its integer multiples. We use the following notation:
\begin{equation}
(p)\equiv\{m p | m\in\mathbb{Z}\}\,.
\end{equation}
An ideal generated by a single element is called a \textit{principal ideal}. We can also consider ideals generated by more than one element:
\begin{equation}
(a,b)\equiv\{m a + n b | m, n \in\mathbb{Z}\}\,.
\end{equation}
Ideals of this kind can be multiplied, with $(a,b)(c,d)=(ac,ad,bc,bd)$. With these concepts in place, we can return to our earlier example. Suppose we wish to factor, not the number $9$, but the ideal $(9)$. Then the factorization in equation \ref{eq:nonunique} can be further refined, by factoring principal ideals into ideals generated by two elements. We have,
\begin{equation}
(9)=(3)\times (3)=(2+\sqrt{-5})\times(2-\sqrt{-5})=(3,1+\sqrt{-5})^2(3,1-\sqrt{-5})^2
\label{eq:unique}
\end{equation}
This factorization is now unique: the ideals $(3,1+\sqrt{-5})$ and $(3,1-\sqrt{-5})$ are not merely irreducible, but prime.

Taking into account some subtleties regarding the unit element, and a generalization to fractional ideals (both of which we will not discuss here), factorization into prime ideals allowed for all remaining symbol letters in these integrals to be represented in terms of a truly multiplicatively independent basis. Writing the symbol in this basis, we found that all roots that were expected to be spurious cancel: the only surviving roots in each integral are those identified as ``physically meaningful''. Out of the original over 2000 symbol letters for each integral we find both integrals can be expressed in a common basis of just 35 symbol letters, leading to symbols that are 5216 and 5245 terms in length, three orders of magnitude smaller than our initial results. Remarkably, we find that in the combination that these two integrals appear in the physical amplitude the remaining algebraic letters actually cancel, and only integer letters remain. This result was later confirmed via other methods~\cite{He:2020vob}.

\section{Parametric Square Roots: Elliptic and Beyond}
\label{sec:4}

Even with the methods of section \ref{sec:2}, some integrals are not linearly reducible. This happens when partial-fractioning or fibration gives rise to an algebraic root in the remaining integration variables that cannot be rationalized. As a square root of a quadratic polynomial can always be rationalized by a rational change of variables, the first nontrivial case involves cubic or quartic polynomials in a single variable. These polynomials define elliptic curves, and there is a growing literature on the Feynman integrals that contain them~\cite{Laporta:2004rb,MullerStach:2012az,brown2011multiple,Bloch:2013tra,Adams:2013kgc,Adams:2014vja,Adams:2015gva,Adams:2015ydq,Adams:2016xah,Adams:2017ejb,Adams:2017tga,Bogner:2017vim,Broedel:2017kkb,Broedel:2017siw,Adams:2018yfj,Broedel:2018iwv,Adams:2018bsn,Broedel:2018qkq,Adams:2018kez,Honemann:2018mrb,Bogner:2019lfa,Broedel:2019hyg,Broedel:2019kmn}. Via direct integration, we found that the two-loop ten-particle N${}^3$MHV amplitude in planar $\mathcal{N}=4$ super Yang-Mills has a supercomponent with this property: linear reducibility is obstructed by an elliptic curve~\cite{Bourjaily:2017bsb}.

In other cases, linear reducibility is obstructed by an algebraic root in more than one variable, for example $\sqrt{Q(x_1,x_2,\ldots)}$. There has been much less progress on Feynman integrals of this kind, but what progress exists has focused on analyzing the geometric properties of the algebraic varieties defined by these roots (for example, by the equation $y^2=Q(x_1,x_2,\ldots)$). In particular, the most productive cases thus far have involved varieties that define Calabi-Yaus~\cite{Groote:2005ay,Brown:2009ta,Brown:2010bw,Bloch:2014qca,Bloch:2016izu,broadhurstprivate,Bourjaily:2018ycu,Bourjaily:2018yfy,Bourjaily:2019hmc,Festi:2018qip,Broedel:2019kmn,Besier:2019hqd,mirrors_and_sunsets,Klemm:2019dbm,Bonisch:2020qmm}.

One way a variety may define a Calabi-Yau is if it can be embedded consistently in a weighted projective space, such that the sum of the coordinate weights is equal to the overall degree of the polynomial~\cite{Hubsch:1992nu}. By ``embedded consistently'' we mean that it must scale uniformly under weighted rescaling of the coordinates. In the following subsections, we will describe several diagrams and classes of diagram that give rise to varieties with this property under direct integration. These examples will have a common structure: they can all be embedded in $k$-dimensional weighted projective space $\mathbb{WP}^{1,\ldots,1,k}$ (where all coordinate weights except one have weight $1$, and the remaining coordinate has weight $k$). The origin of this property will be clear for the first class of diagrams we discuss, and more mysterious for the second.

\subsection{Scalar Marginal Integrals}

For our first set of examples, we consider scalar diagrams, writing them in the well-known Symanzik representation. For an $L$-loop Feynman diagram $\mathcal{I}$ with $E$ internal edges $\{e_i\}$ with masses $m_i$ in $D$ dimensions, we write:
\begin{equation}
\mathcal{I}=\Gamma(E-L D/2)\int_{x_i\geq0}[d^{E-1}x_i]\frac{\mathfrak{U}^{E-(L+1)D/2}}{\mathfrak{F}^{E-L D/2}}\,,
\end{equation}
where the graph polynomials $\mathfrak{U}$ and $\mathfrak{F}$ are defined by:
\begin{equation}
\mathfrak{U}\equiv\sum_{\{T\}\in\mathfrak{T}_1}\prod_{e_i\notin T}x_i,\quad
\mathfrak{F}\equiv\Bigg[\sum_{{\{T_1,T_2\}\in\mathfrak{T}_2}}s_{T_1}\Big(\prod_{{e_i\notin T_1\cup T_2}}x_i\Big)\Bigg]+\mathfrak{U}\sum_{e_i}x_im_i^2\,.
\end{equation}
The $\mathfrak{U}$ polynomial is a sum over spanning trees $\mathfrak{T}_1$ of the graph, while the $\mathfrak{F}$ polynomial includes a sum over disconnected pairs of trees that together span the graph (denoted $\mathfrak{T}_2$). $s_{T_1}$ is the square of the sum of momenta flowing in to tree $T_1$, while $[d^{E-1}x_i]$ denotes projective integration over the $E$ variables $\{x_i\}$.

There are two cases where this representation simplifies dramatically, making it easier to probe its geometry. In the first, consider a case where $E=LD/2$. The function $\Gamma(E-L D/2)$ diverges in this case, but if there are no subdivergences then the rest of the integral can be convergent in integer dimensions:
\begin{equation}
\bar{\mathcal{I}}=\int_{x_i\geq0}[d^{E-1}x_i]\frac{1}{\mathfrak{U}^{D/2}}\,.
\end{equation}
Integrals of this form have no kinematic dependence, so they are simply numbers. They are referred to in the mathematical literature as Feynman periods. These diagrams give rise to Calabi-Yaus upon direct integration~\cite{Brown:2009ta}.

Next, consider a case where $E=(L+1)D/2$. We considered cases of this form in ref.~\cite{Bourjaily:2018yfy}, this subsection reviews our discussion there. Here $\Gamma(E-L D/2)=\Gamma(D/2)$ is finite. Provided the integral is otherwise convergent we refer to this class as marginally convergent, or ``marginal''. In integer dimensions we find:
\begin{equation}
\mathcal{I}=\Gamma(D/2)\int_{x_i\geq0}[d^{E-1}x_i]\frac{1}{\mathfrak{F}^{D/2}}\,.
\end{equation}
Because of the kinematic dependence of $\mathfrak{F}$ this is now not merely a number, but a function.

In two dimensions, scalar marginal integrals consist of the comparatively well-studied higher-loop sunrise diagrams~\cite{Groote:2005ay,Bloch:2014qca,Bloch:2016izu,broadhurstprivate,Broedel:2019kmn,mirrors_and_sunsets,Klemm:2019dbm,Bonisch:2020qmm}. In higher dimensions there are many more such diagrams.

All marginal integrals share common features, which ensure that if linear reducibility is obstructed the obstruction will define a Calabi-Yau. The Symanzik representation of these integrals depends on only the $\mathfrak{F}$ polynomial. This polynomial is homogeneous and has degree $L+1$, so the denominator in the Symanzik representation $\mathfrak{F}^{D/2}$ has degree $(L+1)D/2=E$, the same as the number of variables. Direct integration preserves this property: each integration will decrease the number of variables and the overall degree of the denominator by one. If at some stage partial-fractioning introduces a square root of a polynomial in the remaining $m$ integration parameters $\sqrt{Q(x_i)}$, the polynomial $Q(x_i)$ will be homogeneous and have overall degree $2m$. The equation $y^2=Q(x_i)$ defines a variety. By assigning weight $m$ to $y$ and weight $1$ to each of the $x_i$ we may consistently embed this variety in a weighted projective space. A quick count shows that the sum of these coordinate weights is equal to the degree of our variety, showing that all such varieties will define Calabi-Yaus.

Specializing to graphs with massless particles in four dimensions, we can further prove a bound on the dimension of these spaces. Starting once again with the Symanzik representation,
\begin{equation}
\mathcal{I}=\int_{x_i\geq0}[d^{2L+1}x_i]\frac{1}{\mathfrak{F}^{2}}\,,
\end{equation}
as $\mathfrak{F}$ is linear in each variable when all propagators are massless we can integrate in any variable. Integrating in $x_j$ and writing $\mathfrak{F}=\mathfrak{F}^{(j)}_0+x_j\mathfrak{F}^{(j)}_1$, we obtain
\begin{equation}
\mathcal{I}=\int_{x_i\geq0}[d^{2L}x_i]\frac{1}{\mathfrak{F}^{(j)}_0\mathfrak{F}^{(j)}_1}\,.
\label{eq:2Lstep}
\end{equation}
Each of $\mathfrak{F}^{(j)}_0$ and $\mathfrak{F}^{(j)}_1$ is separately linear in each remaining integration variable, so we can once again integrate in any remaining $x_k$. Writing $\mathfrak{F}^{(j)}_i=\mathfrak{F}^{(j,k)}_{i,0}+x_k\mathfrak{F}^{(j,k)}_{i,1}$, we have
\begin{equation}
\mathcal{I}=\int_{x_i\geq0}[d^{2L-1}x_i]\frac{\log\left(\mathfrak{F}^{(j,k)}_{0,0}\mathfrak{F}^{(j,k)}_{1,1}\right)-\log\left(\mathfrak{F}^{(j,k)}_{0,1}\mathfrak{F}^{(j,k)}_{1,0}\right)}{\mathfrak{F}^{(j,k)}_{0,0}\mathfrak{F}^{(j,k)}_{1,1}-\mathfrak{F}^{(j,k)}_{0,1}\mathfrak{F}^{(j,k)}_{1,0}}\,.
\end{equation}

The denominator of the integrand is now at most quadratic in each variable, while the arguments of the logs are products of polynomials which are linear in each variable. We are thus able to integrate once more, but this time potentially at a cost of introducing a square root of a polynomial in the remaining $2L-1$ variables. If this polynomial is irreducibly quartic or cubic in each remaining integration parameter then the root cannot be rationalized by a rational change of variables. This is thus the first integration step at which direct integration can potentially be obstructed, in cases when there is no integration order that avoids an irreducibly quartic or cubic root. As such, the varieties that characterize this obstruction represent the highest-dimension Calabi-Yaus that can occur for this class of diagrams, demonstrating that Calabi-Yau dimension is bounded with loop order, at a maximum of $2L-2$.

\begin{figure}[t]
\sidecaption[t]
\includegraphics[scale=1.]{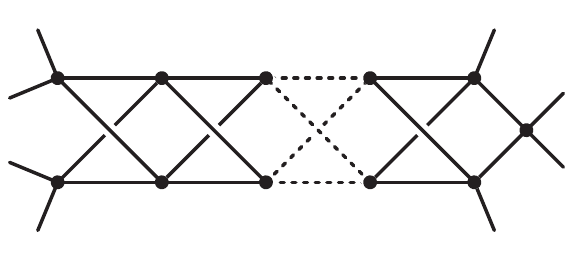}
\caption{The tardigrade diagrams, defined at even loops}
\label{fig:tardigrade}       
\end{figure}

\begin{figure}[b]
\sidecaption
\includegraphics[scale=1.]{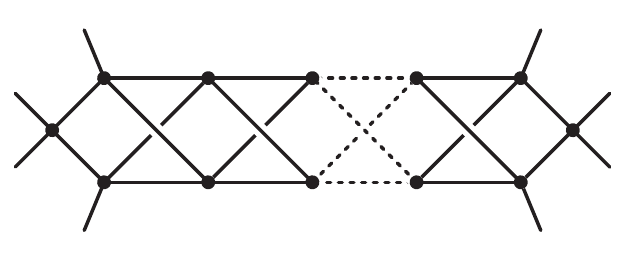}
\caption{The paramecium diagrams, defined at odd loops}
\label{fig:paramecium}       
\end{figure}

\begin{figure}[t]
\sidecaption[t]
\includegraphics[scale=1.]{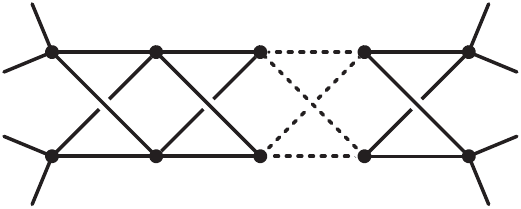}
\caption{The amoeba diagrams, defined at odd loops $\geq 5$}
\label{fig:amoeba}       
\end{figure}

It turns out that this bound is saturated. There are Feynman diagrams at each loop order for which every integration order is obstructed, and the obstruction defines a Calabi-Yau of dimension $2L-2$. We characterize three infinite families of such diagrams. For even loops we find what we refer to as the tardigrade diagrams, depicted in Fig.~\ref{fig:tardigrade}. For odd loops we find two infinite families, which we refer to as paramecia and amoebas, depicted in Fig.~\ref{fig:paramecium} and Fig.~\ref{fig:amoeba} respectively. The amoeba diagrams are not maximally obstructed in this fashion at three loops, but otherwise saturate the bound at each higher order.

We have investigated a broader set of marginal diagrams in four dimensions. We find as the loop order increases, the majority of diagrams have maximal dimension obstructions of this kind.

\subsection{More General Examples}

The previous argument clarifies why Calabi-Yaus appear during direct integration of marginal diagrams. We have also found several examples of non-marginal diagrams that also give rise to Calabi-Yaus.  

In each case discussed here, our starting point will be a $2L$-fold integral over a rational function. While such a representation is easy to obtain from the Symanzik form for marginal integrals (see equation~(\ref{eq:2Lstep}) where this is manifest), for non-marginal integrals it is easier to find these representations via loop-by-loop parametrization (see section~\ref{subsec:loopbyloop}). Heuristically, we believe that starting with a $2L$-fold will ensure that no integration is in any sense spurious, in analogy with the polylogarithmic case of transcendental weight $2L$ at $L$ loops. In particular, since the integrals we consider occur in planar $\mathcal{N}=4$ super Yang-Mills we expect uniform transcendentality: the integral should require exactly $2L$ parameters to express the integrand rationally, and no fewer.

The marginal integrals which were maximally obstructed manifested their obstructions after three integrations, at which point their integrand contained a dilogarithm. The integrals considered here will typically be less obstructed than this, involving Calabi-Yaus of dimension lower than $2L-2$ at $L$ loops. As such, it is in many cases quite cumbersome to perform a full direct integration up to the point it becomes obstructed, particularly for the higher-loop cases. As a proxy for this integration we instead typically took maximal codimension residues, which in rough terms probes whether the integrands can be iteratively partial-fractioned in the integration variables, but may not be sensitive to whether the resulting polylogarithms can at each stage be written in an appropriate fibration basis.

\begin{figure}[b]
\sidecaption
\includegraphics[scale=1.]{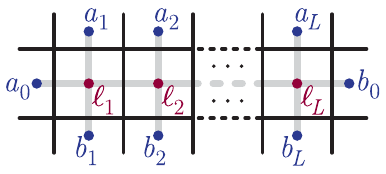}
\caption{The traintrack diagrams. These diagrams are planar, and we label here their dual coordinates.}
\label{fig:traintrack}       
\end{figure}

The first non-marginal diagrams which we found to give rise to Calabi-Yaus were the traintrack diagrams, higher-loop analogues to the elliptic double-box~\cite{Bourjaily:2018ycu}. Depicted in Fig.~\ref{fig:traintrack}, these diagrams, much like the higher-loop sunrise diagrams, increase in Calabi-Yau dimension at each loop order, with dimension $L-1$ at $L$ loops. In ref.~\cite{Bourjaily:2019hmc} we showed that the Calabi-Yau arising from the three-loop traintrack can be written, much like the marginal integrals, as a variety embedded in $\mathbb{WP}^{1,\ldots,1,k}$ (where there $k=3$). Later, ref.~\cite{Vergu:2020uur} analyzed the leading singularity structure of these integrals to all orders in twistor space, finding Calabi-Yau geometry at each order.

\begin{figure}[t]
\sidecaption[t]
\includegraphics[scale=1.]{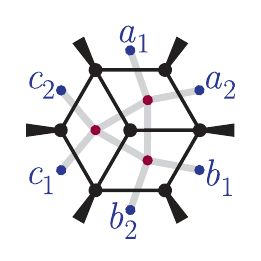}
\caption{The three-loop wheel diagram. This diagram is planar, and we label here its dual coordinates.}
\label{fig:wheel}       
\end{figure}

The three-loop wheel diagram (depicted in Fig.~\ref{fig:wheel}) was also analyzed in ref.~\cite{Bourjaily:2019hmc}, and gives rise to a variety which can be embedded in $\mathbb{WP}^{1,1,1,1,4}$, corresponding to a Calabi-Yau threefold. In this case the embedding is slightly more subtle to derive, involving a deprojectivization and a particular choice of reprojectivization. 

There thus appears to be a common structure in each Calabi-Yau that has been observed in a Feynman diagram in the literature: all cases can be described with varieties embedded in $\mathbb{WP}^{1,\ldots,1,k}$. In ref.~\cite{Bourjaily:2019hmc} we went into some detail analyzing the properties of a generic Calabi-Yau defined in this space, including Hodge numbers and Euler characteristics. It is an open question whether this structure is universal, and if so what it can teach us about these diagrams.

\section{Conclusions}

We reviewed a variety of applications of direct hyperlogarithmic integration in the context of multi-leg scattering amplitudes. We presented tricks that allow one to avoid certain obstructions, but we also highlighted the structure of the obstructions that remain once these tricks are employed: obstructions which in a surprisingly varied set of cases define Calabi-Yaus.

It would be extremely interesting to go beyond the tips and tricks discussed here, and find a systematic algorithm that can find a hyperlogarithmic expression for any Feynman integral for which such an expression exists. One possibility is that such a method might arise from a motivic understanding of these integrals~\cite{Brown:2020rda}.

Along related lines, it will be important to understand which aspects of the Calabi-Yau geometries characterized in this work are universal, and which are specific to particular representations. The results of ref.~\cite{Vergu:2020uur} suggest that such common features exist, but it still may be the case that one can express a single integral with multiple geometries (see for example the role of isogeny in ref.~\cite{Bogner:2019lfa}), or that the obstructions found here need to be augmented  with other information for a full characterization.

Finally, it is worth investigating under what conditions Calabi-Yaus can arise from Feynman integrals. As the work reviewed here shows, they are more common than one might naively assume.

\input{references}

\end{document}

%% file: references.tex
%
%
%

%% file: vonhippel.bbl
\begin{thebibliography}{99.}%
\bibitem{Brown:2009ta}
F.~C.~S.~Brown,
[arXiv:0910.0114 [math.AG]].

\bibitem{Panzer:2015ida}
E.~Panzer,
doi:10.18452/17157
[arXiv:1506.07243 [math-ph]].

\bibitem{Panzer:2014caa}
E.~Panzer,
Comput. Phys. Commun. \textbf{188} (2015), 148-166
doi:10.1016/j.cpc.2014.10.019
[arXiv:1403.3385 [hep-th]].

\bibitem{Bogner:2014mha}
C.~Bogner and F.~Brown,
Commun. Num. Theor. Phys. \textbf{09} (2015), 189-238
doi:10.4310/CNTP.2015.v9.n1.a3
[arXiv:1408.1862 [hep-th]].

\bibitem{Kotikov:1990kg}
A.~V.~Kotikov,
Phys. Lett. B \textbf{254} (1991), 158-164
doi:10.1016/0370-2693(91)90413-K

\bibitem{Remiddi:1997ny}
E.~Remiddi,
Nuovo Cim. A \textbf{110} (1997), 1435-1452
[arXiv:hep-th/9711188 [hep-th]].

\bibitem{MullerStach:2012mp}
S.~M\"uller-Stach, S.~Weinzierl and R.~Zayadeh,
Commun. Math. Phys. \textbf{326} (2014), 237-249
doi:10.1007/s00220-013-1838-3
[arXiv:1212.4389 [hep-ph]].

\bibitem{Henn:2013pwa}
J.~M.~Henn,
Phys. Rev. Lett. \textbf{110} (2013), 251601
doi:10.1103/PhysRevLett.110.251601
[arXiv:1304.1806 [hep-th]].

\bibitem{Lee:2014ioa}
R.~N.~Lee,
JHEP \textbf{04} (2015), 108
doi:10.1007/JHEP04(2015)108
[arXiv:1411.0911 [hep-ph]].

\bibitem{Meyer:2016slj}
C.~Meyer,
JHEP \textbf{04} (2017), 006
doi:10.1007/JHEP04(2017)006
[arXiv:1611.01087 [hep-ph]].

\bibitem{Gluza:2007rt}
J.~Gluza, K.~Kajda and T.~Riemann,
Comput. Phys. Commun. \textbf{177} (2007), 879-893
doi:10.1016/j.cpc.2007.07.001
[arXiv:0704.2423 [hep-ph]].

\bibitem{Blumlein:2014maa}
J.~Bl\"umlein, I.~Dubovyk, J.~Gluza, M.~Ochman, C.~G.~Raab, T.~Riemann and C.~Schneider,
PoS \textbf{LL2014} (2014), 052
doi:10.22323/1.211.0052
[arXiv:1407.7832 [hep-ph]].

\bibitem{Ochman:2015fho}
M.~Ochman and T.~Riemann,
Acta Phys. Polon. B \textbf{46} (2015) no.11, 2117
doi:10.5506/APhysPolB.46.2117
[arXiv:1511.01323 [hep-ph]].

\bibitem{Bern:2005iz}
Z.~Bern, L.~J.~Dixon and V.~A.~Smirnov,
Phys. Rev. D \textbf{72} (2005), 085001
doi:10.1103/PhysRevD.72.085001
[arXiv:hep-th/0505205 [hep-th]].

\bibitem{Bourjaily:2017bsb}
J.~L.~Bourjaily, A.~J.~McLeod, M.~Spradlin, M.~von Hippel and M.~Wilhelm,
Phys. Rev. Lett. \textbf{120} (2018) no.12, 121603
doi:10.1103/PhysRevLett.120.121603
[arXiv:1712.02785 [hep-th]].

\bibitem{Bourjaily:2018ycu}
J.~L.~Bourjaily, Y.~H.~He, A.~J.~Mcleod, M.~Von Hippel and M.~Wilhelm,
Phys. Rev. Lett. \textbf{121} (2018) no.7, 071603
doi:10.1103/PhysRevLett.121.071603
[arXiv:1805.09326 [hep-th]].

\bibitem{Bourjaily:2018aeq}
J.~L.~Bourjaily, A.~J.~McLeod, M.~von Hippel and M.~Wilhelm,
JHEP \textbf{08} (2018), 184
doi:10.1007/JHEP08(2018)184
[arXiv:1805.10281 [hep-th]].

\bibitem{Bourjaily:2018yfy}
J.~L.~Bourjaily, A.~J.~McLeod, M.~von Hippel and M.~Wilhelm,
Phys. Rev. Lett. \textbf{122} (2019) no.3, 031601
doi:10.1103/PhysRevLett.122.031601
[arXiv:1810.07689 [hep-th]].

\bibitem{Bourjaily:2019hmc}
J.~L.~Bourjaily, A.~J.~McLeod, C.~Vergu, M.~Volk, M.~Von Hippel and M.~Wilhelm,
JHEP \textbf{01} (2020), 078
doi:10.1007/JHEP01(2020)078
[arXiv:1910.01534 [hep-th]].

\bibitem{Bourjaily:2019igt}
J.~L.~Bourjaily, A.~J.~McLeod, C.~Vergu, M.~Volk, M.~Von Hippel and M.~Wilhelm,
JHEP \textbf{02} (2020), 025
doi:10.1007/JHEP02(2020)025
[arXiv:1910.14224 [hep-th]].

\bibitem{Bourjaily:2019vby}
J.~L.~Bourjaily, M.~Volk and M.~Von Hippel,
JHEP \textbf{02} (2020), 095
doi:10.1007/JHEP02(2020)095
[arXiv:1912.05690 [hep-th]].

\bibitem{Brown:2008um}
F.~Brown,
Commun. Math. Phys. \textbf{287} (2009), 925-958
doi:10.1007/s00220-009-0740-5
[arXiv:0804.1660 [math.AG]].

\bibitem{Bogner:2013tia}
C.~Bogner and M.~L\"uders,
Contemp. Math. \textbf{648} (2015), 11-28
[arXiv:1302.6215 [hep-ph]].

\bibitem{Besier:2018jen}
M.~Besier, D.~Van Straten and S.~Weinzierl,
Commun. Num. Theor. Phys. \textbf{13} (2019), 253-297
doi:10.4310/CNTP.2019.v13.n2.a1
[arXiv:1809.10983 [hep-th]].

\bibitem{Frellesvig:2017aai}
H.~Frellesvig and C.~G.~Papadopoulos,
JHEP \textbf{04} (2017), 083
doi:10.1007/JHEP04(2017)083
[arXiv:1701.07356 [hep-ph]].

\bibitem{Bourjaily:2019jrk}
J.~L.~Bourjaily, F.~Dulat and E.~Panzer,
Nucl. Phys. B \textbf{942} (2019), 251-302
doi:10.1016/j.nuclphysb.2019.03.022
[arXiv:1901.02887 [hep-th]].

\bibitem{Hodges:2009hk}
A.~Hodges,
JHEP \textbf{05} (2013), 135
doi:10.1007/JHEP05(2013)135
[arXiv:0905.1473 [hep-th]].

\bibitem{Caron-Huot:2018dsv}
S.~Caron-Huot, L.~J.~Dixon, M.~von Hippel, A.~J.~McLeod and G.~Papathanasiou,
JHEP \textbf{07} (2018), 170
doi:10.1007/JHEP07(2018)170
[arXiv:1806.01361 [hep-th]].

\bibitem{Landau:1959fi}
L.~D.~Landau,
Nucl. Phys. \textbf{13} (1960) no.1, 181-192
doi:10.1016/B978-0-08-010586-4.50103-6

\bibitem{Hodges:boxInt}
Andrew~P. Hodges.
\newblock {Crossing and Twistor Diagrams}.
\newblock {\em Twistor Newsletter}, 5:4, 1977.


\bibitem{Hodges:2010kq}
A.~Hodges,
JHEP \textbf{08} (2013), 051
doi:10.1007/JHEP08(2013)051
[arXiv:1004.3323 [hep-th]].

\bibitem{Mason:2010pg}
L.~Mason and D.~Skinner,
J. Phys. A \textbf{44} (2011), 135401
doi:10.1088/1751-8113/44/13/135401
[arXiv:1004.3498 [hep-th]].

\bibitem{Brown:2009qja}
F.~C.~S.~Brown,
Annales Sci. Ecole Norm. Sup. \textbf{42} (2009), 371
[arXiv:math/0606419 [math.AG]].

\bibitem{Goncharov:2010jf}
A.~B.~Goncharov, M.~Spradlin, C.~Vergu and A.~Volovich,
Phys. Rev. Lett. \textbf{105} (2010), 151605
doi:10.1103/PhysRevLett.105.151605
[arXiv:1006.5703 [hep-th]].

\bibitem{sagemath}
{The Sage Developers}.
\newblock {\em {{\tt {S}ageMath}, the {S}age {M}athematics {S}oftware {S}ystem
  ({V}ersion 8.8)}}, 2019.
\newblock Available at: {\url{www.sagemath.org}}.

\bibitem{PARI2}
{The PARI~Group}, Univ. Bordeaux.
\newblock {\em {\texttt{PARI}/\texttt{GP} version {\tt 2.11.0}}}, 2018.
\newblock Available at: \url{pari.math.u-bordeaux.fr}.


\bibitem{He:2020vob}
S.~He, Z.~Li and C.~Zhang,
[arXiv:2009.11471 [hep-th]].

\bibitem{Laporta:2004rb}
S.~Laporta and E.~Remiddi,
Nucl. Phys. B \textbf{704} (2005), 349-386
doi:10.1016/j.nuclphysb.2004.10.044
[arXiv:hep-ph/0406160 [hep-ph]].

\bibitem{MullerStach:2012az}
S.~Muller-Stach, S.~Weinzierl and R.~Zayadeh,
PoS \textbf{LL2012} (2012), 005
doi:10.22323/1.151.0005
[arXiv:1209.3714 [hep-ph]].

\bibitem{brown2011multiple}
F.~Brown and A.~Levin,
\newblock {\em arXiv preprint arXiv:1110.6917}, 2011.


\bibitem{Bloch:2013tra}
S.~Bloch and P.~Vanhove,
J. Number Theor. \textbf{148} (2015), 328-364
doi:10.1016/j.jnt.2014.09.032
[arXiv:1309.5865 [hep-th]].

\bibitem{Adams:2013kgc}
L.~Adams, C.~Bogner and S.~Weinzierl,
J. Math. Phys. \textbf{54} (2013), 052303
doi:10.1063/1.4804996
[arXiv:1302.7004 [hep-ph]].

\bibitem{Adams:2014vja}
L.~Adams, C.~Bogner and S.~Weinzierl,
J. Math. Phys. \textbf{55} (2014) no.10, 102301
doi:10.1063/1.4896563
[arXiv:1405.5640 [hep-ph]].

\bibitem{Adams:2015gva}
L.~Adams, C.~Bogner and S.~Weinzierl,
J. Math. Phys. \textbf{56} (2015) no.7, 072303
doi:10.1063/1.4926985
[arXiv:1504.03255 [hep-ph]].

\bibitem{Adams:2015ydq}
L.~Adams, C.~Bogner and S.~Weinzierl,
J. Math. Phys. \textbf{57} (2016) no.3, 032304
doi:10.1063/1.4944722
[arXiv:1512.05630 [hep-ph]].

\bibitem{Adams:2016xah}
L.~Adams, C.~Bogner, A.~Schweitzer and S.~Weinzierl,
J. Math. Phys. \textbf{57} (2016) no.12, 122302
doi:10.1063/1.4969060
[arXiv:1607.01571 [hep-ph]].

\bibitem{Adams:2017ejb}
L.~Adams and S.~Weinzierl,
Commun. Num. Theor. Phys. \textbf{12} (2018), 193-251
doi:10.4310/CNTP.2018.v12.n2.a1
[arXiv:1704.08895 [hep-ph]].

\bibitem{Adams:2017tga}
L.~Adams, E.~Chaubey and S.~Weinzierl,
Phys. Rev. Lett. \textbf{118} (2017) no.14, 141602
doi:10.1103/PhysRevLett.118.141602
[arXiv:1702.04279 [hep-ph]].

\bibitem{Bogner:2017vim}
C.~Bogner, A.~Schweitzer and S.~Weinzierl,
Nucl. Phys. B \textbf{922} (2017), 528-550
doi:10.1016/j.nuclphysb.2017.07.008
[arXiv:1705.08952 [hep-ph]].

\bibitem{Broedel:2017kkb}
J.~Broedel, C.~Duhr, F.~Dulat and L.~Tancredi,
JHEP \textbf{05} (2018), 093
doi:10.1007/JHEP05(2018)093
[arXiv:1712.07089 [hep-th]].

\bibitem{Broedel:2017siw}
J.~Broedel, C.~Duhr, F.~Dulat and L.~Tancredi,
Phys. Rev. D \textbf{97} (2018) no.11, 116009
doi:10.1103/PhysRevD.97.116009
[arXiv:1712.07095 [hep-ph]].

\bibitem{Adams:2018yfj}
L.~Adams and S.~Weinzierl,
Phys. Lett. B \textbf{781} (2018), 270-278
doi:10.1016/j.physletb.2018.04.002
[arXiv:1802.05020 [hep-ph]].

\bibitem{Broedel:2018iwv}
J.~Broedel, C.~Duhr, F.~Dulat, B.~Penante and L.~Tancredi,
JHEP \textbf{08} (2018), 014
doi:10.1007/JHEP08(2018)014
[arXiv:1803.10256 [hep-th]].

\bibitem{Adams:2018bsn}
L.~Adams, E.~Chaubey and S.~Weinzierl,
Phys. Rev. Lett. \textbf{121} (2018) no.14, 142001
doi:10.1103/PhysRevLett.121.142001
[arXiv:1804.11144 [hep-ph]].

\bibitem{Broedel:2018qkq}
J.~Broedel, C.~Duhr, F.~Dulat, B.~Penante and L.~Tancredi,
JHEP \textbf{01} (2019), 023
doi:10.1007/JHEP01(2019)023
[arXiv:1809.10698 [hep-th]].

\bibitem{Adams:2018kez}
L.~Adams, E.~Chaubey and S.~Weinzierl,
JHEP \textbf{10} (2018), 206
doi:10.1007/JHEP10(2018)206
[arXiv:1806.04981 [hep-ph]].

\bibitem{Honemann:2018mrb}
I.~H\"onemann, K.~Tempest and S.~Weinzierl,
Phys. Rev. D \textbf{98} (2018) no.11, 113008
doi:10.1103/PhysRevD.98.113008
[arXiv:1811.09308 [hep-ph]].

\bibitem{Bogner:2019lfa}
C.~Bogner, S.~M\"uller-Stach and S.~Weinzierl,
Nucl. Phys. B \textbf{954} (2020), 114991
doi:10.1016/j.nuclphysb.2020.114991
[arXiv:1907.01251 [hep-th]].

\bibitem{Broedel:2019hyg}
J.~Broedel, C.~Duhr, F.~Dulat, B.~Penante and L.~Tancredi,
JHEP \textbf{05} (2019), 120
doi:10.1007/JHEP05(2019)120
[arXiv:1902.09971 [hep-ph]].

\bibitem{Broedel:2019kmn}
J.~Broedel, C.~Duhr, F.~Dulat, R.~Marzucca, B.~Penante and L.~Tancredi,
JHEP \textbf{09} (2019), 112
doi:10.1007/JHEP09(2019)112
[arXiv:1907.03787 [hep-th]].

\bibitem{Groote:2005ay}
S.~Groote, J.~G.~Korner and A.~A.~Pivovarov,
Annals Phys. \textbf{322} (2007), 2374-2445
doi:10.1016/j.aop.2006.11.001
[arXiv:hep-ph/0506286 [hep-ph]].

\bibitem{Brown:2010bw}
F.~Brown and O.~Schnetz,
Duke Math. J. \textbf{161} (2012) no.10, 1817-1862
doi:10.1215/00127094-1644201
[arXiv:1006.4064 [math.AG]].

\bibitem{Bloch:2014qca}
S.~Bloch, M.~Kerr and P.~Vanhove,
Compos. Math. \textbf{151} (2015) no.12, 2329-2375
doi:10.1112/S0010437X15007472
[arXiv:1406.2664 [hep-th]].

\bibitem{Bloch:2016izu}
S.~Bloch, M.~Kerr and P.~Vanhove,
Adv. Theor. Math. Phys. \textbf{21} (2017), 1373-1453
doi:10.4310/ATMP.2017.v21.n6.a1
[arXiv:1601.08181 [hep-th]].

\bibitem{broadhurstprivate}
D.~Broadhurst,
\newblock \!\!, private communication.


\bibitem{Festi:2018qip}
D.~Festi and D.~van Straten,
Commun. Num. Theor. Phys. \textbf{13} (2019), 463-485
doi:10.4310/CNTP.2019.v13.n2.a4
[arXiv:1809.04970 [math.AG]].

\bibitem{Besier:2019hqd}
M.~Besier, D.~Festi, M.~Harrison and B.~Naskrecki,
Commun. Num. Theor. Phys. \textbf{14} (2020) no.4, 863-911
doi:10.4310/CNTP.2020.v14.n4.a4
[arXiv:1908.01079 [math.AG]].

\bibitem{mirrors_and_sunsets}
C.~Doran, A.~Novoseltsev, and P.~Vanhove,
\newblock To appear.


\bibitem{Klemm:2019dbm}
A.~Klemm, C.~Nega and R.~Safari,
JHEP \textbf{04} (2020), 088
doi:10.1007/JHEP04(2020)088
[arXiv:1912.06201 [hep-th]].

\bibitem{Bonisch:2020qmm}
K.~B\"onisch, F.~Fischbach, A.~Klemm, C.~Nega and R.~Safari,
[arXiv:2008.10574 [hep-th]].

\bibitem{Hubsch:1992nu}
T.~Hubsch,
\textit{Calabi-Yau manifolds: A Bestiary for physicists}, (World Scientific, 1992)

\bibitem{Vergu:2020uur}
C.~Vergu and M.~Volk,
JHEP \textbf{07} (2020), 160
doi:10.1007/JHEP07(2020)160
[arXiv:2005.08771 [hep-th]].

\bibitem{Brown:2020rda}
F.~Brown and C.~Duhr,
[arXiv:2006.09413 [hep-th]].





%
\end{thebibliography}
